\documentclass[achemso,amsmath,amssymb,amsfonts]{revtex4-1}
\usepackage{graphicx}  
\usepackage{bm}        
\usepackage{amssymb}   
\usepackage{natbib}
\usepackage{amsmath}
\usepackage[colorlinks=true,citecolor=red,linkcolor=blue]{hyperref}
\begin{document}
\title{Linear and nonlinear optics of hybrid plasmon-exciton nanomaterials in the presence of overlapping resonances}

\author{Maxim Sukharev}
\email{maxim.sukharev@asu.edu}
\affiliation{Science and Mathematics Faculty, College of Letters and Sciences, Arizona State University, Mesa, Arizona 85212}

\author{Paul N. Day}
\affiliation{Materials and Manufacturing Directorate, Air Force Research Laboratory, Wright-Patterson Air Force Base, Ohio 45433}

\author{Ruth Pachter}
\email{ruth.pachter@us.af.mil}
\affiliation{Materials and Manufacturing Directorate, Air Force Research Laboratory, Wright-Patterson Air Force Base, Ohio 45433}

\date{\today}
\begin{abstract}
We consider a hybrid plasmon-exciton system comprised of a resonant molecular subsystem and three Au wires supporting a dipole mode which can be coupled to a dark mode in controllable fashion by variation of a symmetry parameter. The physics of such a system under strong coupling conditions is examined in detail. It is shown that if two wires supporting the dark mode are covered with molecular layers the system exhibits four resonant modes for a strong coupling regime due to asymmetry and lifted degeneracy of the molecular state in this case, while upon having molecular aggregates covering the top wire with dipolar mode, three resonant modes appear. Pump-probe simulations are performed to scrutinize the quantum dynamics and find possible ways to control plasmon-exciton materials. It is demonstrated that one can design hybrid nanomaterials with highly pronounced Fano-type resonances when excited by femtosecond lasers.
\end{abstract}
\maketitle

\section{Introduction}
 \label{introduction}
Understanding plasmon-exciton or so-called plexitonic coupling \cite{doi:10.1021/nl200579f} in hybrid plasmonic nanostructures is important for tuning the optical response, e.g. for applications in nonlinear optics \cite{doi:10.1021/cm5039914}, organic solar cells \cite{In:2015aa}, or organic light-emitting diodes \cite{Yadav:2014aa}. In developing such nanostructures, it is important to consider strong coupling phenomena. Indeed, experimentally, Rabi oscillations were observed in J-aggregate/metal nanostructures \cite{DeLacy:2015aa,Vasa:2013aa,PhysRevLett.114.036802,doi:10.1021/nl4014887}. For example, in the work by Schlather et al. \cite{doi:10.1021/nl4014887}, gold nanodisk dimers were utilized, with J-aggregates formed from monomers of a cyanine dye. Cynanine dyes were also used in the work by DeLacy et al. \cite{DeLacy:2015aa} with silver nanoplatelets. Limiting cases of Rabi oscillations/Fano resonances were identified, where the plasmon resonance has an extremely narrow/large linewidth \cite{Faucheaux:2014aa}.

Inducing a Rabi splitting, for example, could be useful in gaining increased tunability in the response. However, such tunability is limited, e.g. in varying the distance between dimers. Here we consider an asymmetric metal plasmonic nanostructure in conjunction with a quantum excitonic subsystem, and particularly the effects of the excitonic subsytem on the coupling between the dipolar and quadrupole modes. We adopted the geometry similar to that described by Gallinet et al. \cite{doi:10.1021/nl303896d} by varying the symmetry parameter. In previous works, we considered plexitonic nanomaterials both in linear \cite{PhysRevLett.109.073002} and nonlinear \cite{doi:10.1021/nn4054528,jcp_chirps14} regimes. It was shown that in addition to commonly observed Rabi splittings, plexitonic systems exhibit collective resonances at high exciton densities. Such resonances correspond to collective electromagnetic modes induced by strong exciton-exciton interactions greatly enhanced by plasmons \cite{PhysRevLett.109.073002}. The existence of plasmon-enhanced collective exciton modes have also been confirmed in core-shell materials \cite{doi:10.1021/ph500032d}. Moreover using the pump-probe technique one can modify optical properties of plexitonic structures via modifications of exciton populations \cite{doi:10.1021/nn4054528} and efficiently control electromagnetic localization both in space and time \cite{jcp_chirps14}.

Using a self-consistent rigorous approach, we show that upon strong coupling in the asymmetric structure similar to the one in Ref. [\onlinecite{doi:10.1021/nl303896d}], and with inclusion of a strongly coupled excitonic subsystem, also varying the position of deposition, interesting phenomena appear that have not been elucidated yet. Such systems improve on the tunability of plexitonic hybrid materials. Furthermore, in performing pump-probe simulations we note appearance of a pronounced Fano lineshape.

\section{Theoretical model}
 \label{model}
A spatiotemporal dynamics of electric, $\vec{E}$, and magnetic, $\vec{H}$, fields is considered classically using the full machinery of the Maxwell equations
   \begin{subequations}
   \label{Maxwell}
    \begin{align}
     & \mu_0 \frac{\partial \vec{H}}{\partial t}  =  -\nabla \times \vec{E} \\
     & \varepsilon_0\frac{\partial \vec{E}}{\partial t}  =  \nabla \times \vec{H} - \vec{J}, 
    \end{align}
  \end{subequations}
 where $\varepsilon_0$ and $\mu_0$ are the permittivity and the permeability of the free space, respectively, and $\vec{J}$ corresponds to either the current density in spatial regions occupied by metal or the macroscopic polarization current, $\vec{J}=\frac{\partial\vec{P}}{\partial t}$, in space with molecules. The equations (\ref{Maxwell}) are integrated using the finite-difference time-domain (FDTD) approach, which implicitly accounts for Gauss's law via Yee's cell thus requiring only two equations to solve.

To account for the material dispersion in a metal the Drude model is implemented. In this model the current density is evaluated according to the following equation \cite{PhysRevB.68.045415}
 \begin{equation}
\label{Drude_J}
 \frac{\partial\vec{J}}{\partial t}+\Gamma\vec{J}=\varepsilon_0\omega_p^2\vec{E},
\end{equation}
where $\Gamma$ is the damping parameter and $\omega_p$ is the bulk plasma frequency. An additional parameter that enters the Drude model is the high-frequency limit of the dielectric constant $\varepsilon_r$. For the range of frequencies considered in this work the following set of parameters was chosen to represent gold: $\varepsilon_r=9.5$, $\omega_p=8.95$ eV, and $\Gamma=0.069$ eV. We note that we tested a more precise description of the dielectric function of gold using the Drude-Lorentz model \cite{Rakic:98}. The results were qualitatively similar to those obtained with the Drude model, i.e. all resonances discussed below were present but their energy positions were slightly different. Moreover the physical nature of all resonances were the same as we checked via calculating corresponding charge distributions. The Drude model however leads to significantly shorter execution times of our codes hence it was used to capture essential physics rather than quantitative features.

The optical response of a molecular aggregate is simulated using rate equations for a two-level system \cite{siegman1986university} driven by a local electric field $\vec{E}$
   \begin{subequations}
   \label{rate_equations}
    \begin{align}
&\frac{dn_1}{dt}-\gamma_{21}n_2=-\frac{1}{\hbar\Omega_0}\vec{E}\cdot\frac{\partial\vec{P}}{\partial t}, \\
&\frac{dn_2}{dt}+\gamma_{21}n_2=\frac{1}{\hbar\Omega_0}\vec{E}\cdot\frac{\partial\vec{P}}{\partial t}, \\
&\frac{\partial^2\vec{P}}{\partial t^2}+(\gamma_{21}+2\gamma_{d})\frac{\partial\vec{P}}{\partial t}+\Omega_0^2\vec{P}=-\sigma(n_2-n1)\vec{E},
    \end{align}
 \end{subequations}
where $n_1$ and $n_2$ describe the populations of the ground and the excited molecular states, respectively, $\vec{P}$ is the macroscopic polarization, $\gamma_{21}$ is the radiationless decay rate of the excited state, $\gamma_{d}$ is the pure dephasing rate, and $\hbar\Omega_0$ is the energy separation of the molecular levels. The coupling constant $\sigma$ can be derived from a simple harmonic oscillator as readily shown in Ref. [\onlinecite{DIET_PRA}]
   \begin{equation}
   \label{constants}
\sigma=\frac{2\Omega_0\mu_{12}}{3\hbar},
  \end{equation}
here $\mu_{12}$ is the transition dipole moment.

The equations (\ref{Maxwell}) and (\ref{rate_equations}) are coupled via polarization current $\vec{J}=\frac{\partial\vec{P}}{\partial t}$ that appears in the Ampere law. The resulting system of equations is solved numerically on a multi-processor computer. We note that in such an approach the static molecule-molecule interactions are neglected thus allowing us to treat each molecule independently. This method however does account for all interactions between molecules which are induced by external EM radiation. The values of the molecular parameters used in this work are: $\mu_{12}=10$ Debye, $\gamma_{21}=6.892\times10^{-4}$ eV (corresponding to $6$ ps), $\gamma_{d}=6.565\times10^{-3}$ eV (corresponding to $630$ fs), and the total number density of molecules is $n_0=4\times10^{25}$ m${^{-3}}$. The thickness of a molecular layer in all simulations below is $10$ nm. Finally we assume that the dielectric host for all structures considered in this work has a dielectric constant $\varepsilon=1.7689$ corresponding to that of water.

The space is discretized in accordance with the FDTD algorithm \cite{taflove2000computational} in three dimensions. The spatial resolution is $\delta x=\delta y=\delta z=1.5$ nm to achieve numerical convergence and avoid staircase effects. To ensure numerical convergence for pump-probe simulations as discussed in details in Section \ref{pump-probe} we repeated several runs with a spatial resolution of $1.0$ nm and $0.8$ nm. The results were nearly identical to those obtained with a $1.5$ nm step.The time step is $\delta t=\delta x/(2c)=2.5$ as, where $c$ is the speed of light in vacuum. Open boundaries are simulated using convolutional perfectly matched layers (CPML) \cite{CPML_paper}. We found that for systems considered here the best results were achieved with $19$ CPML layers.

The excitation of a system is carried out using the total field/scattered field approach \cite{taflove2000computational}, which allows one to inject a plane wave into a simulation domain. For calculations of linear spectra we employ a short-pulse method \cite{PhysRevA.84.043802} with a time pulse envelope, $f\left(t\right)$, written in the form of the Blackman-Harris window
   \begin{equation}
   \label{Blackman-Harris}
f\left(t\right)=\sum_{n=-0}^{3}a_n \text{cos}\left(\frac{2\pi n t}{\tau}\right)
  \end{equation}
for a pulse with a duration $\tau$, where $a_0=0.353222222$, $a_1=-0.488$, $a_2=0.145$, and $a_3=-0.010222222$. The same time envelop was used to simulate a femtosecond pump.

\begin{figure}
\begin{center}
\includegraphics[width=0.9\textwidth]{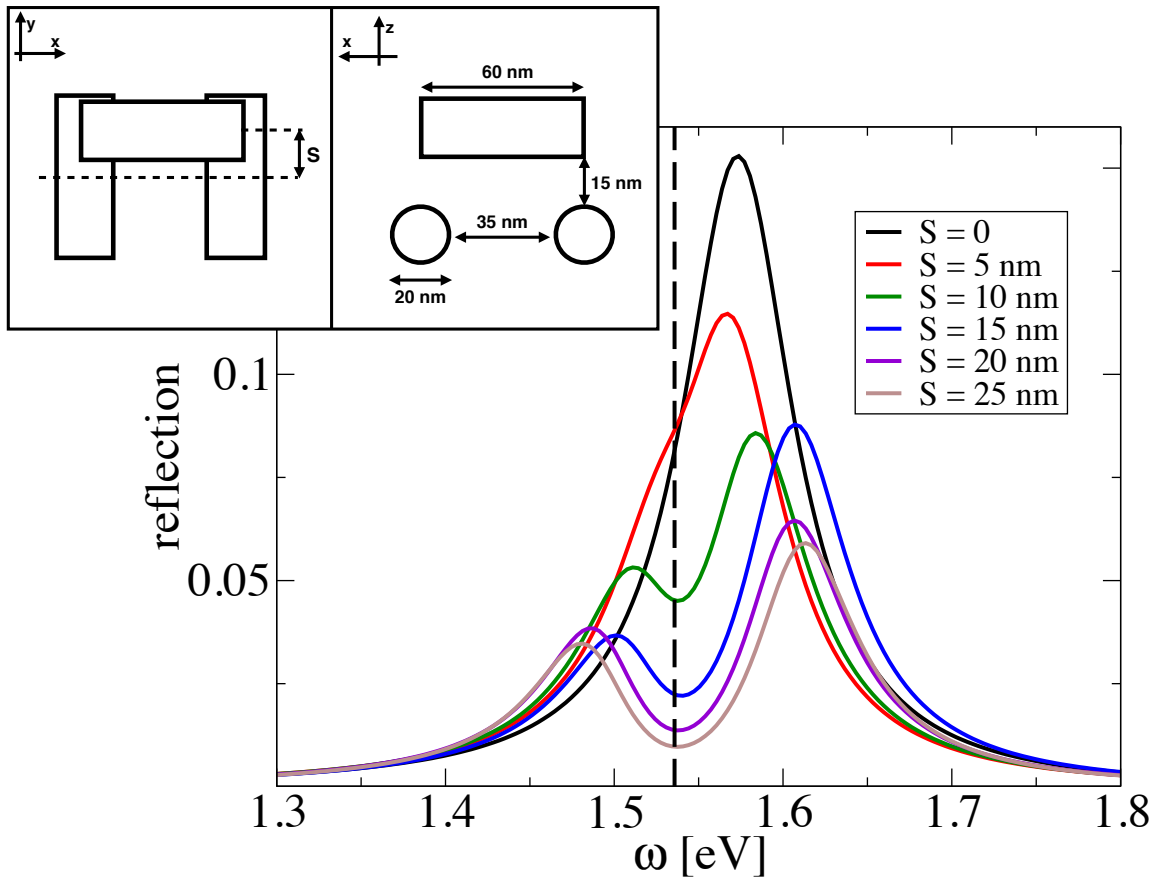}
\caption{\label{fig1} (Color online) The inset shows the arrangement of gold nanowires in a three-dimensional structure. The system is excited by a plane wave polarized along the $x$-axis and propagating along the negative $z$ direction. The fixed geometrical parameters such as the length of each wire, etc. are indicated in the inset. The main panel shows reflection spectra for the system as a function of the incident photon energy at different positions of the upper wire as indicated in the main panel legend. The vertical dashed line shows the position of the quadrupole mode at $1.53574$ eV.
}
\end{center}
\end{figure}

To test our numerical approach we qualitatively compare results of our simulations with data from Ref [\onlinecite{doi:10.1021/nl303896d}]. The structure considered in this work is schematically depicted in the inset of Fig. \ref{fig1}. To characterize this system we place a pointwise detector above the upper wire and compute the EM energy flux in the positive $z$ direction. The subsequent normalization of these calculations to the energy flux of the incident wave results in reflection as plotted in the main panel of Fig. \ref{fig1}.

The upper wire acts as a dipole antenna coupling to incident radiation. The bottom two wires support a dark mode which can not be directly excited by the external field. The dipole antenna however depending on the asymmetry parameter $S$ (see the inset in Fig. \ref{fig1}) allows one to harvest the dark mode. The main panel shows the reflection calculated at a point near the system. It is seen from this figure that the more asymmetric the structure is the stronger the coupling between the dipole mode of the upper wire and the dark mode of the two parallel wires is. This observation is in a good agreement with main results of Ref [\onlinecite{doi:10.1021/nl303896d}]. We note that the dipole mode for the upper wire is at $1.57364$ eV while in Ref [\onlinecite{doi:10.1021/nl303896d}] it is near $1.6$ eV. The discrepancy comes from the fact that in our simulations the length of each wire was set to $60$ nm (as opposed to $100$ nm in Ref [\onlinecite{doi:10.1021/nl303896d}]) leading to a higher energy of dipole oscillations. Although a slight offset of the dipole and dark modes results in lower overall coupling since both resonances are sufficiently broad the strong coupling regime between these modes is clearly observed.

\section{Results and discussion}
\label{results}
The main goal of this work is to explore how the presence of resonant molecular aggregates influences coupling between bright and dark modes in the system shown in Fig. \ref{fig1}. We investigate two scenarios of a deposition of molecular aggregates. First (\textit{geometry A}) is to cover the upper wire with a thin layer of resonant molecules to alter the dipole mode which in turn changes the indirect coupling between the incident field and the dark mode of the system. The second scenario (\textit{geometry B}) is to place molecules on a surface of two parallel lower wires in order to see if a resonant molecular system can more efficiently change the coupling by interacting with the dark mode directly. Moreover the spatial proximity of resonant molecular layers in the case of geometry B should lead to a degenerate molecular state and hence more than three modes in spectra at the strong coupling conditions.

\subsection{Linear optical response}
\label{linear}

\begin{figure}
\begin{center}
\includegraphics[width=0.9\textwidth]{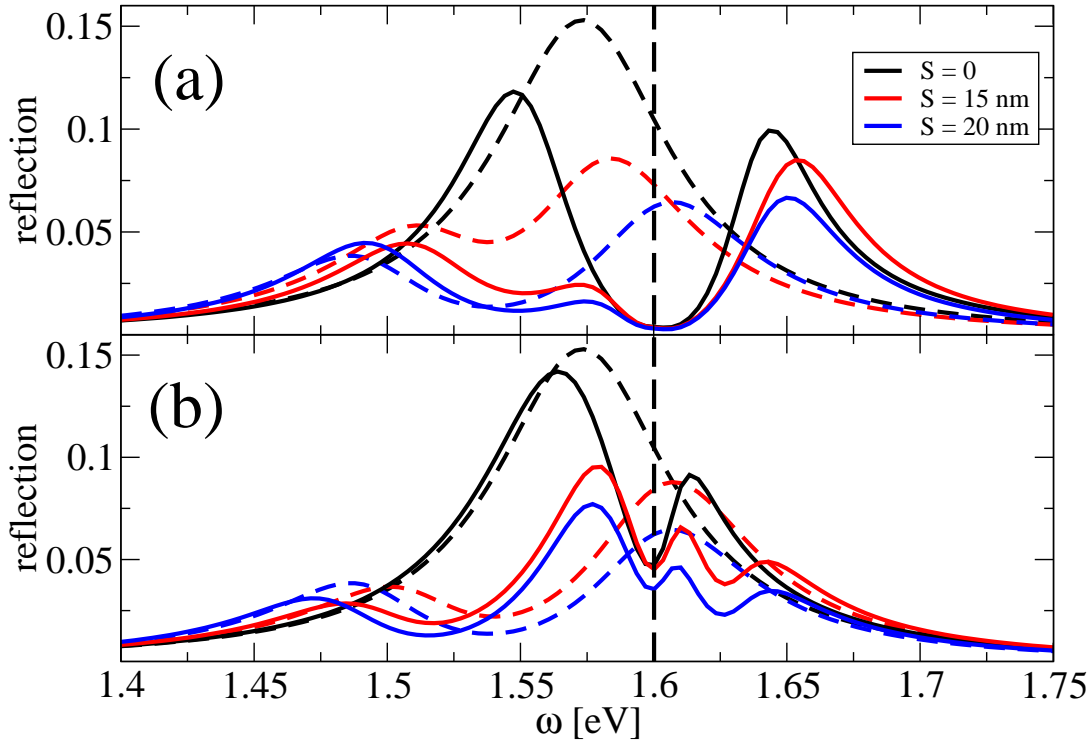}
\caption{\label{fig2} (Color online) Reflection spectra for two scenarios of a deposition of molecular aggregates. Panel (a) shows results as solid lines for the molecular layer covering the upper wire (see the inset in Fig. \ref{fig1} for details). Panel (b) shows data as solid lines obtained with two lower wires covered by resonant molecules. In both scenarios the thickness of a molecular layer is $10$ nm with all molecules resonant at $\Omega_0=1.6$ eV as indicated by the vertical dashed line. Dashed lines in both panels correspond to reflection spectra without molecular aggregates.
}
\end{center}
\end{figure}
First we examine spectral features of the hybrid system (plasmons+excitons) in the linear regime calculating reflection as a function of the incident photon energy. The linear regime in our calculations corresponds to molecular excited state populations always much smaller than $1$.

Results for both molecular coverages are shown in Fig. \ref{fig2} for the case of a molecular resonance centered at $\Omega_0=1.6$ eV. We note several interesting observations. The arrangement corresponding to geometry A shows a clear signature of the strong coupling between molecules and the dipole mode due to strong local field enhancement as it was initially anticipated. For the symmetric orientation of the upper wire, $S=0$, one can see that the Rabi splitting of the dipole resonance is about $97$ meV. For asymmetric cases the splitting increases reaching the value of $158$ meV at  $S=20$ nm. It is also clear from Fig. \ref{fig2}a that asymmetric arrangements of wires have three distinct resonant modes corresponding to two obvious hybrid exciton-plasmon modes due to the coupling between molecules and the dipole mode of the upper wire and the dark mode as expected. 

In comparing our results (Fig. \ref{fig2}) with the electromagnetic transparency window and energy storage results for the asymmetric structure in Ref. [\onlinecite{doi:10.1021/nl303896d}], we note a reduction in intensity for both the A and B coating suppositions in our system for $S=0$, as expected, although larger than the reduction in intensity of more than 50$\%$ that was calculated for the strong coupling regime for the system in Ref. [\onlinecite{doi:10.1021/nl303896d}]. This has been rationalized in terms of the radiative and nonradiative decay contributions. It is interesting to note that in the case of $S=0$, where the strong coupling with the molecular layer induces a Rabi splitting, subsystem B demonstrates a blue-shift for the first peak as compared to subsystem A and the splitting is less pronounced, as is expected. This introduces another parameter that could enable modulation of the reflectance in these geometries.

One may anticipate that for geometry B two lower wires covered with resonant molecules would exhibit additional interaction: molecular layer on wire $1$ with molecular layer on wire $2$. If the coupling between these layers and either the dipole mode or the dark mode is strong the spectral response may show four resonances. The results for geometry B are shown in Fig. \ref{fig2}b. For the symmetric case, $S=0$, we again observe the Rabi splitting but this time it is noticeably smaller, $50$ meV, compared to the previous case. This can be explained by the fact that the molecular layers now cover two lower wires and its EM coupling to the dipole mode supported by the upper wire is lower. For asymmetric arrangements there are four clearly distinctive resonances observed. This indicates the strong coupling between all four subsystems: the dipole mode of the upper wire, the quadrupole mode of the two parallel lower wires, the molecular aggregate on wire $1$, and the molecular aggregate on wire $2$. The two last modes are degenerate with the degeneracy lifted by their coupling with plasmon states of the three interacting wires.

\begin{figure}
\begin{center}
\includegraphics[width=0.9\textwidth]{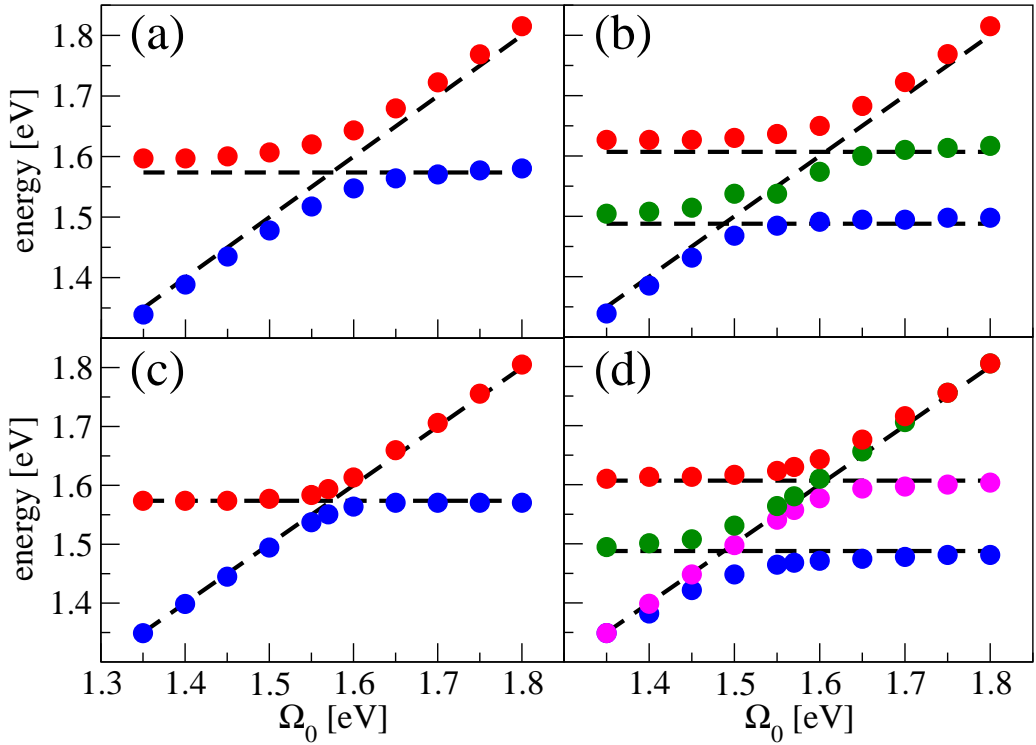}
\caption{\label{fig3} (Color online) Avoided crossings for two sets of geometries. Panels (a) and (b) show energy positions of all resolved resonances in reflection spectra as functions of the molecular transition energy $\Omega_0$ for geometry A (molecular layer covers the upper wire only). Panels (c) and (d) show energies of all resonances for geometry B (molecular layer covers two lower parallel wires). The asymmetry parameter $S=0$ for panels (a) and (c), and $S=20$ nm for panels (b) and (d). Note that each resonance is represented by circles with different colors. Horizontal dashed lines in each panel show corresponding non-interacting plasmon modes. For the symmetric case $S=0$ there is only one plasmon state while at $S=20$ nm there are two states as discussed in Fig. \ref{fig1}.
}
\end{center}
\end{figure}

To further demonstrate the physics of such a system we performed a series of simulations varying the molecular resonant energy $\Omega_0$ for both geometries A and B and extracting energies of each mode for a given molecular transition energy. The results are presented in Fig. \ref{fig3}. Both geometries A and B with the symmetric arrangement $S=0$ (Figs. \ref{fig3}a and \ref{fig3}c) exhibit a single avoided crossing indicating the coupling between molecular aggregates and the dipole mode supported by the upper wire. Once the symmetry is broken ($S=20$ nm, Figs. \ref{fig3}b and \ref{fig3}d) a set of multiple avoided crossings appears. In the case of geometry A, where a single molecular state is coupled to both dipole-dark plasmon states as Fig. \ref{fig3}b demonstrates with clearly distinct avoided crossings. Geometry B with a non-zero asymmetry parameter $S$ exhibits more complex spectra (Fig. \ref{fig3}d). At low and high molecular transition energies both molecular layers show a single resonance due to weak interaction between them. However when the molecular energy of both layers approaches the region of the strong coupling with hybrid plasmon states the degeneracy is removed and spectra show four distinct resonances. It is interesting to note that when the molecular energy is about $1.57$ eV (close to the dark plasmon mode) one of the molecular states is very narrow. The proximity of a narrow and a wide resonance could be used to coherently control EM energy distribution in such a system if properly altered by a strong external laser pulse inducing the Fano-type interference.

\subsection{Pump-probe simulations}
\label{pump-probe}
\begin{figure}
\begin{center}
\includegraphics[width=0.9\textwidth]{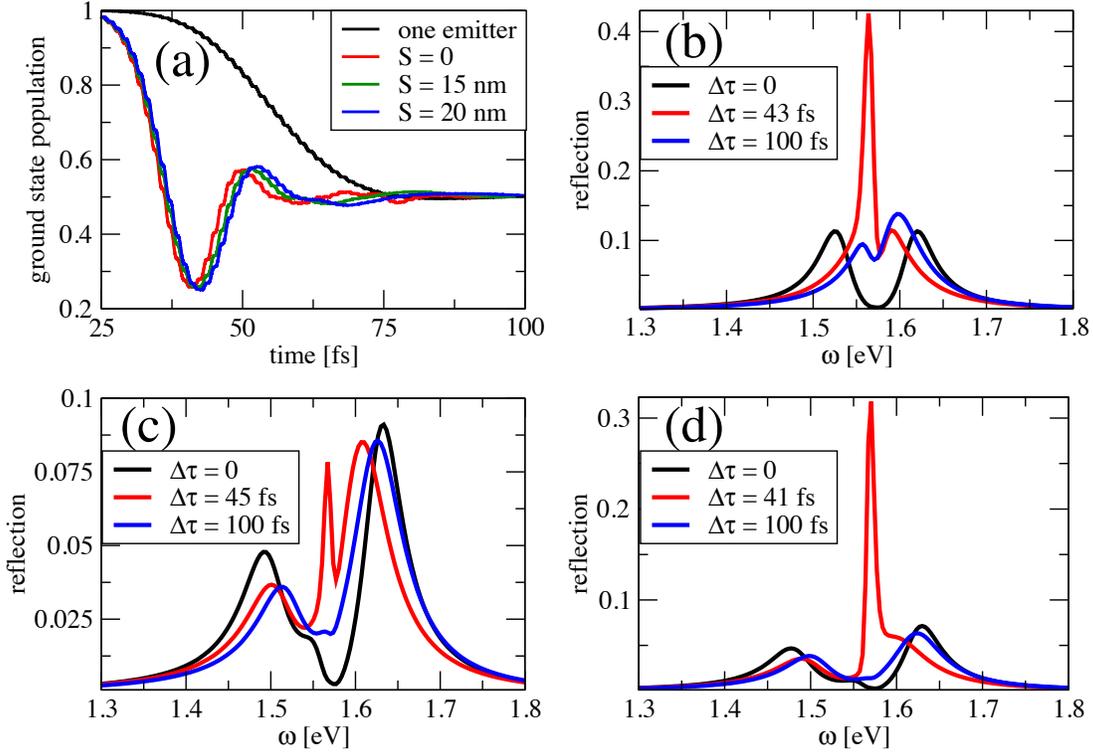}
\caption{\label{fig4} (Color online) Pump-probe simulations for geometry A. Panel (a) shows time dynamics of the ground state population. Black lines show data for a single molecule in vacuum subject to the pump pulse. Red, green, and blue lines show ensemble average populations of the ground state as a function of time for $S=0$, $S=15$ nm, and $S=20$ nm, respectively. Panel (b) shows reflection as a function of the probe photon energy with $S=0$ at three pump-probe time delays: black line is for $\Delta\tau=0$ (system is not altered by pump yet), red line is for $\Delta\tau=43$ fs (panel (b)), $\Delta\tau=45$ fs (panel (c)), $\Delta\tau=41$ fs (panel (d)), and blue line is for $\Delta\tau=100$ fs. Panels (c) and (d) show data similar to that presented in panel (b) but for $S=15$ nm and $S=20$ nm, respectively. The pump pulse parameters are: pulse duration is $100$ fs, peak amplitude is $E_0=2.7483\times10^8$ V$/$m, central frequency is $1.57$ eV. The molecular transition energy is $1.57$ eV.
}
\end{center}
\end{figure}

We consider a pump-probe experiment during which a system under consideration is subject to strong external laser radiation. After some time delay, $\Delta\tau$, a low intense probe pulse arrives and "measures" the corresponding \textit{linear} response of the system altered by the pump. Numerical aspects of such simulations for hybrid plasmon-exciton systems can be found in Ref. [\onlinecite{doi:10.1021/nn4054528}]. In all simulations presented below we define the pump-probe time delay such that $\Delta\tau=0$ corresponds to the probe arriving just before the pump. 

First we examine geometry A pumping it with a $100$ fs laser pulse with an amplitude corresponding to a $\pi/2$-pulse (i.e. pumping a half of molecular population into the excited state). The results of the simulations are presented in Fig. \ref{fig4}. The panel (a) shows the time dynamics of the average populations of the molecular ground state compared to a single molecule case in vacuum. One can see that the dynamics of the hybrid system is significantly different from a single emitter case - it is noticeably faster exhibiting several Rabi cycles with diminishing amplitude. We note that observed faster dynamics can be explained by local field enhancement due to the excitation of a corresponding plasmon mode. Higher local field increases the pump pulse area leading to more Rabi cycles. Another important aspect to point out is a quick decay of the amplitude of Rabi oscillations. Upon careful examination of spatial distributions of excited molecules one can conclude that the excitation travels fast spreading over the entire molecular layer. Since the layer is not uniformly exposed to the pump this excitation is highly inhomogeneous leading to a quick dephasing of macroscopic polarization currents \cite{Vasa:2013aa}. Another evidence to support this explanation is that the Rabi splitting at the end of the pump is considerably lower compared to its unperturbed value (for $S=0$ the splitting is $94$ meV for the unperturbed spectrum and is only $42$ meV at the end of the pump). The excitation spreads quickly over the entire molecular layer thus leading to a high dephasing which in turn lowers the coupling with the hybrid plasmon mode and hence affects the Rabi splitting.

\begin{figure}
\begin{center}
\includegraphics[width=0.9\textwidth]{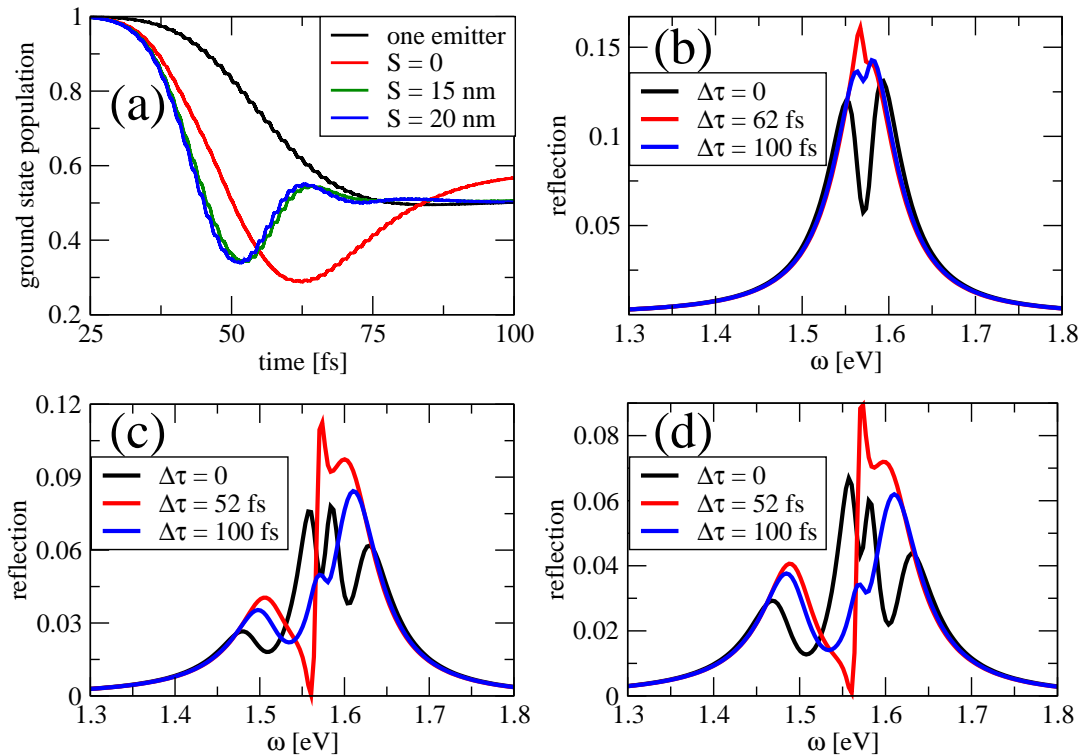}
\caption{\label{fig5} (Color online) Same as in Fig. \ref{fig4} but for geometry B. Parameters of the molecular system and the pump are the same.
}
\end{center}
\end{figure}
To scrutinize the pump dynamics further we probe the perturbed system at different times as shown in Figs. \ref{fig4}b,  \ref{fig4}c, and  \ref{fig4}d. The system undergoes a strong transition near $40-50$ fs as seen in Fig. \ref{fig4}a as a deep minimum of the ground state population, which reaches almost $20\%$. Probing the system at these times reveals that the molecular aggregate transitions into excited state. The reflection spectra calculated around $\Delta\tau=40$ fs exhibit a narrow resonance at the molecular transition energy $1.57$ eV as the system releases the energy it stored due to the pump. Probed at later times the system shows less and less transitions occurring at $1.57$ eV as the excitation quickly dephases as seen from blue lines in Figs. \ref{fig4}b,  \ref{fig4}c, and  \ref{fig4}d.

\begin{figure}
\begin{center}
\includegraphics[width=0.9\textwidth]{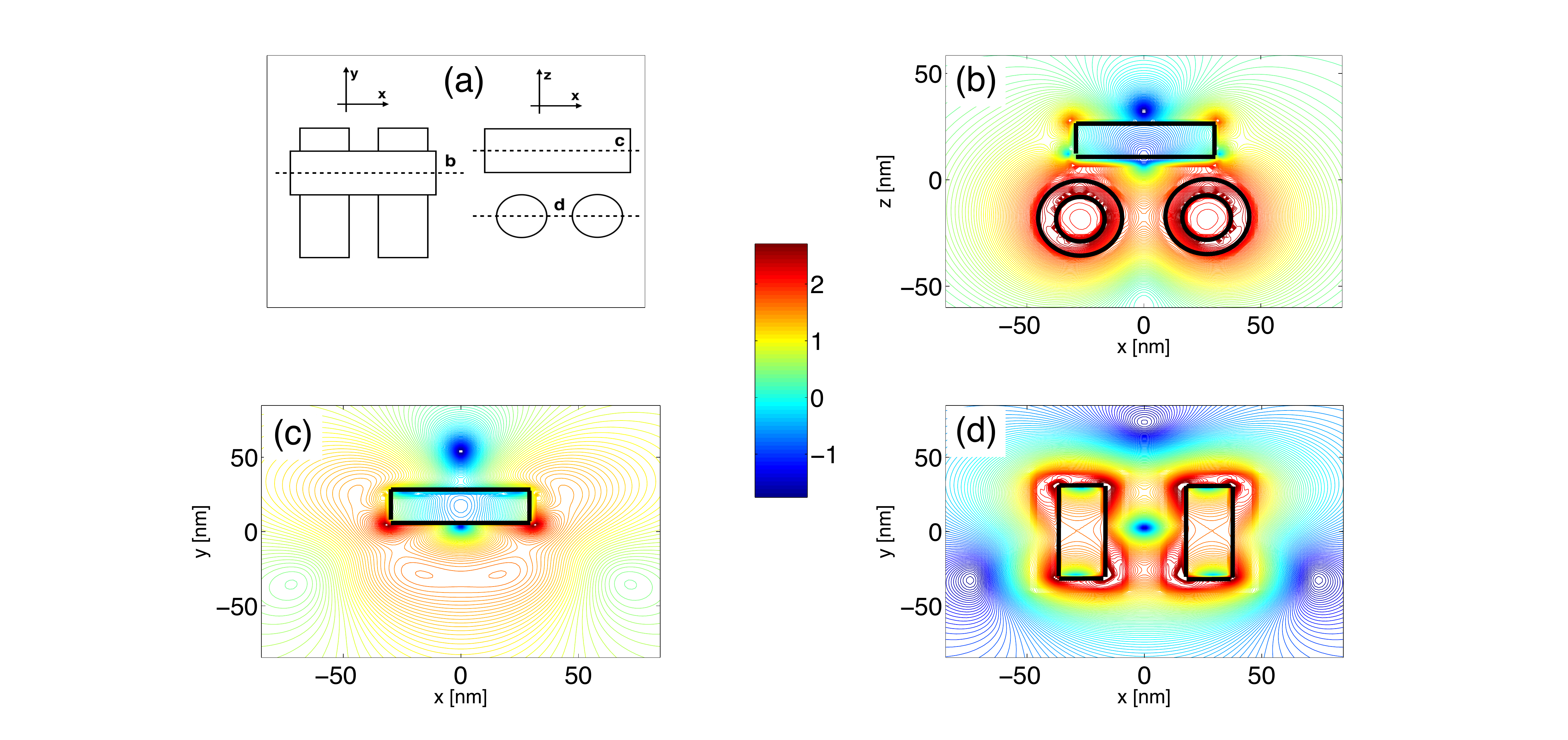}
\caption{\label{fig6} (Color online) Steady-state electromagnetic intensity distributions calculated for $\omega=1.56049$ eV corresponding to the minimum in reflection spectrum shown in Fig. \ref{fig5}c, red line. Panel (a) shows schematics of two-dimensional cuts with each dashed line indicating the corresponding plane, at which intensity is plotted in other panels. Panels (b) - (d) show intensity distributions are plotted in logarithmic scale normalized with respect to the incident intensity, i.e. units of intensity enhancement. Black contour lines depict actual wires along with molecular layers. The incident field is polarized along $x$ and propagates in the negative $z$-direction.
}
\end{center}
\end{figure}

Completely different physics is seen when examining geometry B as shown in Fig. \ref{fig5}. Firstly the quantum dynamics of the molecular subsystem (Fig. \ref{fig5}a) is noticeably slower compared to its counterpart (Fig. \ref{fig4}a). The latter shows that all arrangements with different asymmetry parameters result in similar time dynamics. On the contrary, the symmetric case with $S=0$ for geometry B is considerably slower than others. Probing the system with $S=0$ at different time delays $\Delta\tau$ shows basically the same trend we noted above, i.e. at the time when the average ground state population has a minimum the reflection spectrum has a sharp resonance indicating the fact that a portion of molecules is in the excited state. It again decays very quickly and relaxes back to the unperturbed spectrum due to quick dephasing. However for all asymmetric cases with $S=15$ nm (Fig. \ref{fig5}c) and $S=20$ nm (Fig. \ref{fig5}d) we observe the strong induced Fano-type resonance with the reflection becoming negligibly small at $1.56$ eV. As we anticipated in Section \ref{linear} while discussing avoided crossings the proximity of sharp and broad overlapping resonances may lead to strong interference resulting in Fano lineshapes. Here the interference is caused by the excitation of one of the hybrid molecular-plasmon states and its strong coupling with another hybrid state. We note that such a highly pronounced Fano lineshape with a deep minimum in the reflection leads to high local field enhancement if being excited at the frequency corresponding to this minimum \cite{Stockman:11}.

To illustrate how the system behaves at frequencies corresponding to very small reflection as seen in Fig. \ref{fig5}c and \ref{fig5}d we calculate steady-state electromagnetic energy distributions as shown in Fig. \ref{fig6}. First the structure is pumped by a femtosecond pulse and then probed with a time delay of $\Delta\tau=52$ fs at the frequency of $\omega=1.56049$ eV corresponding to the minimum in reflection (see Fig. \ref{fig5}c, red line). It should be noted that simulations of transmission yield spectra similar to reflection with the same positions of minima and maxima. This suggests that the structure does not scatter the incident radiation and acts as a very efficient absorber. Intensity distributions support this hypothesis demonstrating high localization of electromagnetic energy in regions occupied by molecules with local intensity enhancement reaching $3$ orders of magnitude. As it was proposed in Ref. [\onlinecite{doi:10.1021/nn4054528}], one can pump a plexitonic structure with a femtosecond pulse and abruptly turn the pump off at a desired time leaving the structure in a given state (as in our case turning the pump off after $52$ fs, for instance). A properly prepared system would act as an efficient nano-absorber for a time corresponding to characteristic relaxation time needed for molecules to decay back to their ground state.

\section{Conclusion}
\label{conclusion}
We scrutinized optical properties of a hybrid plasmon-exciton system comprised of a resonant molecular subsystem and three Au wires supporting a dipole mode which can be coupled to a dark mode in controllable fashion by variation of a symmetry parameter. It was shown that if two wires supporting the dark mode are covered with molecular layers the system exhibits four resonant modes for a strong coupling due to asymmetry and lifted degeneracy of the molecular state in this case, while upon having molecular aggregates covering the top wire with dipolar mode, three resonant modes appear. Furthermore we used pump-probe simulations to study the quantum dynamics and possible ways to control plasmon-exciton materials. It was demonstrated that one can design hybrid nanomaterials with highly pronounced Fano-type resonances using femtosecond lasers. We showed that one can use ultra-short laser pulses to design very efficient nano-absorbers. The femtosecond optical engineering of plexitonic systems can be used to create materials with desired properties and functionality.

\section*{Acknowledgements}
  The authors acknowledge support from the Air Force Office of Scientific Research (Summer Faculty Fellowship 2013).

\end{document}